\documentclass{aa}

\usepackage{amsmath}
\usepackage{graphicx}
\usepackage{float} 
\usepackage{txfonts}
\usepackage{stfloats}
\usepackage{placeins}
\usepackage[colorlinks, citecolor=blue, linkcolor=blue]{hyperref}
\usepackage{subcaption} 
\usepackage[utf8]{inputenc}
\usepackage{booktabs}

\providecommand{\gaia}{\textsl{Gaia }}
\providecommand{\gaianospace}{\textit{Gaia}}
\providecommand{\gaianir}{\textsl{GaiaNIR }}
\providecommand{\gaianirnospace}{\textit{GaiaNIR}}

\providecommand{\kmsnospace}{km$~$s$^{-1}$}

\providecommand{\xgboost}{\texttt{XGBoost} }

\newcommand{\kmskpc}{km$~$s$^{-1}~$kpc$^{-1}$ }
\newcommand{\kmskpcnospace}{km$~$s$^{-1}~$kpc$^{-1}$}

\providecommand{\masyr}{mas$~$yr$^{-1}$ }
\providecommand{\masyrnospace}{mas$~$yr$^{-1}$}

\defcitealias{andrae23}{A23}
\defcitealias{drimmel22}{GC23}

\begin{document}

\title{From \textit{Gaia} to \textit{GaiaNIR}:\\II. A new view of the Milky Way bar}

\author{Ó. Jiménez-Arranz\inst{1}$^,$\thanks{Email: oscar.jimenez\_arranz@fysik.lu.se}
   \and M. Schölch\inst{2,3,4}
   \and S. Khanna\inst{5}
   \and L. Chemin\inst{6}
   \and M. Romero-Gómez\inst{2,3,4}
   \and J.~A.~S. Hunt\inst{7}
   \and \\ R. Drimmel\inst{5}
   \and E. Poggio\inst{5}
   \and D. Hobbs\inst{1}
   \and P.~J. McMillan\inst{8}
   \and I. Henum\inst{1}
   \and R.~P. Church\inst{1}
}

\institute{
{Lund Observatory, Division of Astrophysics, Department of Physics, Lund University, Box 118, SE-22100, Lund, Sweden}
\and
{Departament de Física Quàntica i Astrofísica (FQA), Universitat de Barcelona (UB), C Martí i Franquès, 1, 08028 Barcelona, Spain}
\and
{Institut de Ciències del Cosmos (ICCUB), Universitat de Barcelona, Martí i Franquès 1, 08028 Barcelona, Spain}
\and
{Institut d'Estudis Espacials de Catalunya (IEEC), c. Esteve Terradas 1, 08860 Castelldefels (Barcelona), Spain}
\and
{INAF – Osservatorio Astrofisico di Torino, via Osservatorio 20, 10025 Pino Torinese (TO), Italy}
\and
{Observatoire Astronomique de Strasbourg, Université de Strasbourg/CNRS, 11 rue de l’Université, 67000 Strasbourg, France}
\and
{School of Mathematics \& Physics, University of Surrey, Stag Hill, Guildford GU2 7XH, UK}
\and
{School of Physics \& Astronomy, University of Leicester, University Road, Leicester, LE1 7RH, UK}
}

\date{Received <date> / Accepted <date>}

\abstract 
{The Milky Way (MW) hosts a central bar whose pattern speed, orientation, and length remain uncertain, largely due to observational biases and selection effects, despite the transformative data provided by the \gaia mission.}
{We aim to reassess the MW bar properties using \gaia DR3, explicitly accounting for incompleteness and astrometric uncertainties, and to quantify the expected improvements from future \gaia DR4, DR5, and \gaianir data.}
{We combine \gaia DR3 RGB samples with line-of-sight velocities and realistic \gaia and \gaianir mock catalogues to characterise observational biases. We then apply standard techniques to infer the bar pattern speed and structural properties, and evaluate their performance for upcoming data releases.}
{Using \gaia DR3 RGB mock catalogues, we find that the bar pattern speed exhibits a systematic offset of $+14.4 \pm 2.3$ \kmskpcnospace. Applying this approach to the data yields $\Omega_p = 43.7 \pm 0.1$ \kmskpcnospace, which we interpret as a conservative upper limit. Correcting for this bias gives $\Omega_p = 29.3 \pm 2.3$ \kmskpcnospace, although this estimate should be treated with caution given the limited number of mock realizations. We also detect bisymmetric perturbations in $v_\phi$ and $\langle |v_R / v_{\rm tot}| \rangle$, with phase angles $\phi_b = 19$–$24^\circ$ in the bar region. Future \gaia data releases, together with \gaianirnospace, are expected to reduce systematic offsets in the pattern speed to $\sim +5$ \kmskpcnospace. In addition, \gaianir will further improve proper motion precision to below $0.001$ \masyr for bright sources and extend the spatial coverage.}
{Our results indicate that current measurements of the MW bar pattern speed are significantly affected by systematics, but that forthcoming \gaia and \gaianir data will substantially improve both accuracy and robustness.}


\maketitle

\section{Introduction}
\label{sec:introduction}

Galactic bars are a ubiquitous structural feature, present in approximately two-thirds of spiral galaxies in the local \citep[e.g.][]{devaucouleurs91,eskridge00,sheth08,aguerri09,masters11,erwin18,lee19} and early Universe \citep[e.g.][]{leconte24,guo25}. From a theoretical viewpoint, their existence in low-density environments indicates that a disc is dynamically cold and self-gravitating enough to permit non-axisymmetric instabilities \citep[e.g.][]{hohl71,kalnajs72,ostriker_peebles73,sellwood_wilkinson93}. Once formed, bars profoundly influence the dynamical evolution of their host galaxies: by acting as both sources and sinks of angular momentum, they facilitate the redistribution of stars, gas, and dark matter within the galactic disc and halo. Through these processes, bars are widely recognized as fundamental drivers of secular evolution in disc galaxies \citep[e.g.][]{lyndenbell72,athanassoula02,athanassoula03,debattista06,sellwood14}.

Our Galaxy, the Milky Way (MW), was shown to host a central bar more than thirty years ago, based on a combination of infrared photometry from COBE and models of gas kinematics in the inner Galaxy \citep[e.g.][]{binney91,blitz_spergel91,weinberg92,weiland94,dwek95}. Since then, a succession of large-scale surveys has drastically improved our view of the Galactic centre. In particular, the \gaia mission \citep{gaiadr2mission,gaiadr2summary,gaiaedr3summary,gaiadr3summary} has revolutionised studies of the MW by providing precise astrometry for almost two billion stars, enabling comprehensive investigations of Galactic structure and dynamics. Despite this progress, observational limitations — such as the use of heterogeneous stellar tracers, along with varying modelling assumptions — limit the accuracy of measurements of the bar’s key parameters. As a result, inferred values of the bar’s length, pattern speed, orientation angle, and density structure still scatter at the $\sim$20\% level \citep[see reviews before and after the \gaia\ mission][]{bland-hawthorn_gerhard16,shen_zheng20,hunt_vasiliev25}.

While \gaia has been transformative, several of its limitations are particularly acute for studies of the MW bar. The bar resides predominantly in the highly extinguished inner disc \citep[e.g.][]{rieke89,scoville03,gao13,haggard24}, where optical observations are strongly suppressed and the \gaia selection function becomes complex and non-uniform \citep[e.g.][]{cantat-gaudin23,castro-ginard23}. Consequently, \gaia samples of bar-tracing populations are incomplete and biased toward lower-extinction windows and the near side of the bar. Moreover, distance estimation in these regions is challenging: parallax uncertainties increase rapidly with distance \citep[e.g.][]{luri18}, extinction-driven systematics affect photometric priors \citep[e.g.][]{bailer-jones21,anders22}, and line-of-sight crowding complicates the interpretation of astrometric solutions \citep[e.g.][]{fabricius21,everall_boubert22}. These issues directly affect measurements of the bar’s three-dimensional structure, including its length and orientation relative to the Sun, as well as its inferred dynamical properties.

Among the bar’s dynamical parameters, the pattern speed $\Omega_p$ is of particular importance. It characterises the approximately rigid rotation of the bar, governs the locations of major resonances in the disc, and provides insight into the bar’s evolutionary state through angular momentum exchange with the bulge and dark matter halo \citep[e.g.][]{sellwood80,2000debattista,athanassoula02,athanassoula03,athanassoula14}. Historically, measurements of bar pattern speeds have relied on spectroscopic observations, most notably through the method introduced by \citet{tw84}, alongside morphological methods based on resonance feature locations \citep[e.g.][]{rautiainen08}. With the advent of the \gaia mission, however, it has become possible to constrain $\Omega_p$ using astrometric data alone for both the MW \citep[e.g.][]{Bovy2019,drimmel22,zhang24} and nearby galaxies such as the LMC \citep{jimenez-arranz24a,araya25}.

Using \gaia data, a broad consensus has emerged in favour of a relatively long ($\sim$4 kpc), slowly rotating MW bar ($\Omega_p \sim 35$–$40$ \kmskpcnospace) whose major axis is oriented at an angle of $\sim$20–30$^\circ$ with respect to the Sun–Galactic-centre line \citep[e.g.][]{portail17,hunt18,Bovy2019,binney20,chiba-schonrich21,drimmel22,clarke-gerhard22,zhang24,horta25,khalil2025}. However, most existing measurements of the MW bar pattern speed \citep[see Fig.~10 of][]{hunt_vasiliev25} are based on samples that are incomplete in the inner Galaxy and rely on distance estimates affected by non-negligible systematics. Since astrometry-based determinations of the bar pattern speed implicitly depend on adequate spatial coverage and reliable distances, the impact of these limitations on current \gaianospace-based constraints—and on the inferred bar parameters more generally—has not yet been fully quantified and merits a careful reassessment.

Overcoming these limitations requires high-precision astrometry in the near-infrared, where extinction effects are dramatically reduced. The proposed \gaianir mission \citep[][Hobbs et al. in prep.]{hobbs21} is designed to provide such capabilities, delivering accurate positions, parallaxes, and proper motions for stellar populations embedded deep within the inner disc. By enabling a more complete and homogeneous sampling of bar-tracing stars across the full three-dimensional volume of the MW bar, \gaianir will allow substantially improved and internally consistent constraints on the bar’s length, orientation, and pattern speed.

In this work, we use two large samples of red giant branch (RGB) stars from \gaia DR3 \citep[][hereafter referred as A23 and GC23, respectively]{andrae23,drimmel22} to investigate the limitations that incompleteness and distance uncertainties impose on current determinations of the MW bar’s properties. By analysing realistic \gaia mock catalogues \citep[as developed in][]{scholch25} and examining the sensitivity of astrometry-based inferences to selection effects and distance systematics, we provide a critical reassessment of previous \gaianospace-era measurements of $\Omega_p$. Beyond revisiting existing results, a central goal of this paper is to assess how near-infrared astrometry will transform studies of the Galactic bar. Using mock catalogues, we quantify how the improved completeness, more reliable distance estimates and better spatial coverage expected from \gaianir will reduce the dominant systematics affecting current measurements. This will enable substantially more accurate and internally consistent constraints on the MW bar’s pattern speed, as well as its length and orientation, thereby offering a clearer and more robust characterisation of the bar’s structural and dynamical state. Overall, this study provides both a critical appraisal of what can presently be inferred about the MW bar from \gaia data and a quantitative benchmark for the improvements anticipated in the future with \gaianirnospace.

The paper is structured as follows. Section~\ref{sec:data} presents the \gaia DR3 RGB dataset and the MW-like simulations used to validate bar characterisation. Section~\ref{sec:methodology} describes the construction of the \gaia DR3 RGB mock catalogues and outlines the methods employed to characterise the MW bar. In Sect.~\ref{sec:results}, we present our main results from the \gaia DR3 RGB data. Section~\ref{sec:discussion} evaluates the robustness of these findings and compares them with previous estimates. Section~\ref{sec:future} explores the prospects for studying the MW bar with upcoming data releases. Finally, Sect.~\ref{sec:conclusions} summarises the study.

\section{Data}
\label{sec:data}

In this work, we make use of two distinct datasets. First, in Sect. \ref{subsec:rgb}, we describe the \gaia DR3 RGB samples used to infer the MW bar properties. Second, in Sect. \ref{subsec:simulations}, we present the MW-like simulations employed to generate the \gaia DR3 and \gaianir RGB mock catalogues later used to validate the results.

\subsection{\gaia DR3 RGB samples}
\label{subsec:rgb}

We use two independent RGB samples to investigate the capabilities and limitations of \gaia DR3 for characterising the MW bar. The primary sample adopted in this work is based on the RGB catalogue compiled by \citetalias{andrae23}. We also consider, for comparison, the RGB sample presented by \citetalias{drimmel22}.

As previously stated, the first sample is drawn from the \gaia DR3 RGB catalogue compiled by \citetalias{andrae23}, which provides stellar parameters inferred from \gaia XP spectra\footnote{The XP spectra consist of two very-low-resolution spectrographs ($R \sim 40$–$150$) on board the \gaia mission, obtained with the BP and RP prisms \citep{carrasco21,deangeli23}. Together, they cover the wavelength range from $\sim350$ to $\sim1000$ nm \citep{montegriffo23}. In the following, we refer to these BP and RP data collectively as XP spectra.}. In that work, the authors trained \xgboost models \citep{chen_guestrin16} to estimate stellar metallicity ([M/H]), effective temperature ($T_{\mathrm{eff}}$), and surface gravity ($\log g$) from XP spectra, using a subset of sources with reliable external stellar parameters. The initial sample corresponds to the vetted subsample of 17,558,141 RGB stars. This subset was constructed to minimize spurious metallicity estimates, particularly among metal-poor stars, by removing contaminants caused by unrecognized hotter but reddened stars (see Fig.~16 of \citetalias{andrae23}). The selection was designed with two main goals: (1) to isolate a bright giant sample with precise and robust [M/H] estimates suitable for Galactic chemodynamical studies, and (2) to limit the temperature range to $T_{\mathrm{eff}} < 5200$~K, which effectively suppresses contamination from hotter stars. We refer the reader to \citetalias{andrae23} for a full description of the model training and validation procedures. In this work, we apply the same quality and spatial criteria used by \citetalias{drimmel22}, namely an astrometric fidelity cut of $f_a > 0.5$ \citep{rybizki22} and a vertical height limit of $|Z| < 1$~kpc to focus on the Galactic disc. After these selections, the \citetalias{andrae23} sample contains 10,517,183 RGB stars with \gaia RVS line-of-sight velocity measurements.

The other RGB sample used in this work was constructed by \citetalias{drimmel22} using stellar parameters from the GSP-Phot module in \gaia DR3 \citep{gsp_phot}. In this case, RGB stars were selected by requiring $3000 < T_{\mathrm{eff}} < 5500$~K and $\log g < 3.0$, yielding an initial sample of about 11.6 million stars \citepalias[see Fig.~5 of][for the corresponding Kiel diagram]{drimmel22}. Applying the same astrometric fidelity and spatial cuts ($f_a > 0.5$, $|Z| < 1$~kpc) resulted in a refined sample of approximately 8.7 million RGB stars, of which about 5.7 million have RVS line-of-sight velocity measurements. 

Within the bar region ($[R_0, R_1] = [1.0, 4.0]$~kpc; see Sect.~\ref{subsec:dehnen}), the \citetalias{andrae23} sample increases the number of RGB stars of \citetalias{drimmel22} by roughly 10\%, from about 705 thousand to more than 785 thousand. Table~\ref{tabl:data} summarises the source counts for the different \gaia DR3 RGB samples.

\begingroup
\setlength{\tabcolsep}{10pt}
\renewcommand{\arraystretch}{1.5}
\begin{table}
\caption{Number of sources in the different \gaia DR3 RGB samples with available line-of-sight velocity information ($N$), with separate counts provided for the bar region ($N_\text{bar}$). The corresponding MW bar pattern speed inferred from the DSS method ($\Omega_\text{p,DSS}$) is also listed.}
\centering
\begin{tabular}{lccc}
\hline \hline
RGB sample      &  $N$   & $N_\text{bar}$  &  $\Omega_\text{p,DSS}$  \\ \hline
\citetalias{andrae23}    &   10~517~183          &   787~580  &    43.7 $\pm$   0.1    \\ 
\citetalias{drimmel22}      &   5~730~578     & 704~663    &   43.7 $\pm$   0.1     \\ \hline
\end{tabular}
\tablefoot{The bar pattern speed is provided in \kmskpcnospace, and the MW bar region is defined by  [$R_0, R_1$] = [$1.0, 4.0$ kpc].}
\label{tabl:data}
\end{table}
\endgroup

To map the RGBs into a 3D Cartesian framework, we transform the \gaia astrometric observables $(\varpi, \alpha, \delta)$ into spatial coordinates. Since our sample extends to large heliocentric distances (i.e. small parallaxes), positional uncertainties are dominated by parallax errors, allowing us to neglect the comparatively minor contributions from angular uncertainties and their correlations. Heliocentric distances were therefore adopted from the photogeometric estimates of \citet{bailer-jones21}, providing more robust distance determinations for the RGB sample than parallax inversion. Using these distances, we derived heliocentric Cartesian coordinates $(x, y, z)$ assuming a standard 3D Euclidean geometry. These were subsequently transformed into Galactocentric coordinates $(X, Y, Z)$ adopting the Sun’s position at $X_\odot=-R_\odot = -8277 \pm 9_{\text{stat}} \pm 30_{\text{sys}}$~pc \citep{gravity22} and, for simplicity, $Z_\odot = 0$, given that the small vertical offset has a negligible effect on the large-scale disc mapping considered here. In this reference system, $Y$ follows the direction of Galactic rotation (clock-wise rotation), and $Z$ points toward the North Galactic Pole. To map the velocities, we combine the spectroscopically measured line-of-sight velocities with the proper motions and the distance estimates. For a comprehensive account of the procedure used to transform observational quantities into six-dimensional phase-space coordinates, we refer the reader to Sect.~3 of \citetalias{drimmel22}.

\begin{figure*}
    \centering
    \includegraphics[width=\textwidth]{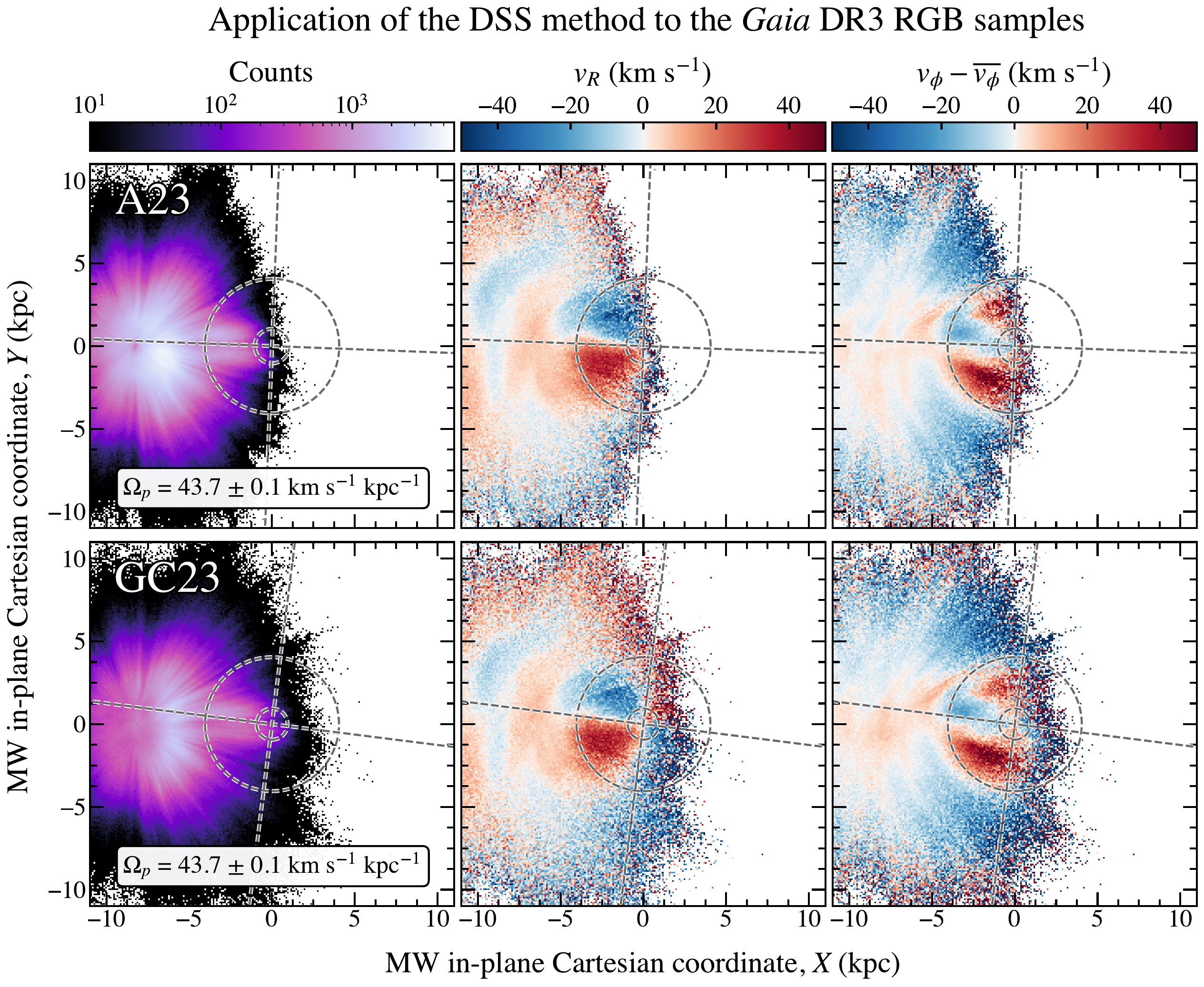}
    \caption{Application of the DSS method to the \citetalias{andrae23} (top) and \citetalias{drimmel22} (bottom) \gaia DR3 RGB samples. Shown are the surface density (left), median radial velocity (centre), and median residual tangential velocity (right) maps as if the MW was seen face-on. The grey dashed circles mark the bar region used in the DSS method, [$R_0, R_1$] = [$1.0, 4.0$ kpc], while the dark grey dashed lines indicate the bar’s minor and major axes as determined by the method. The bar pattern speed $\Omega_p$ derived from the DSS method is indicated at the bottom of the left panels. In this reference frame, the Sun is located at $(X, Y) \approx (-8.3, 0)$~kpc, while the Galactic center is at $(X, Y) = (0, 0)$~kpc.}
    \label{fig:data_maps}
\end{figure*}

Figure \ref{fig:data_maps} displays the surface density (left), median radial velocity (centre), and median residual tangential velocity (right) maps as if the MW was seen face-on. Here, the residual tangential velocity corresponds to the tangential velocity after subtracting the underlying rotation curve. The enhanced coverage in \citetalias{andrae23} relative to \citetalias{drimmel22} stems mainly from improved sampling of the Solar neighbourhood, with a comparatively smaller impact on the bar region -- an increase of approximately 10\%. In both samples, a pronounced drop in density is observed around $X \sim 0$, with the \citetalias{drimmel22} sample extending slightly farther in $X$ than the \citetalias{andrae23} sample. This feature is attributable to observational limitations caused by extinction in the inner Galaxy. In both velocity maps, the bar’s quadrupole pattern is partially discernible.

\subsection{MW-like simulations}
\label{subsec:simulations}

We evaluate how accurately the Galactic bar’s properties can be recovered from \gaia DR3 RGB samples using four MW-like simulations, accounting for incompleteness and astrometric uncertainties. To evaluate the robustness of our results, we use two test-particle simulations, TP40 and TP42, and two self-consistent pure $N$-body simulations, N35 and N49 -- where the number denotes the true bar pattern speed, in \kmskpcnospace, obtained via finite differences.

For the test-particle simulations (TP40 and TP42), we adopted the same initial conditions, Galactic potential, and integration procedure as described in \citet{RomeroGomez2015}, and summarised in Appendix~C of \citetalias{drimmel22}, which are briefly outlined here. The initial positions and velocities were drawn from a disc density distribution following a Miyamoto–Nagai potential \citep{Miyamoto1975}, with a scale height of $h_z = 300$~pc and a radial velocity dispersion of $\sigma_U = 30.3$~km~s$^{-1}$, representative of red clump (RC) stars. The initial conditions were first integrated in the axisymmetric potential of \citet{Allen1991} for 10~Gyr. The Galactic bar potential was then introduced adiabatically over four bar rotations, followed by an additional four bar rotations to allow the system to reach statistical equilibrium. The bar is modelled as the superposition of two aligned Ferrers ellipsoids \citep{Ferrers1877}: one representing the triaxial bulge (semi-major axis of 3.13~kpc), and the other the long thin bar (semi-major axis of 4.5~kpc). The bar rotates as a rigid body with a constant pattern speed $\Omega_p$ of 40 and 42 \kmskpc for the TP40 and TP42 simulations, respectively.

For the pure $N$-body simulations (N35 and N49), the bar slows over time through interaction with the live dark matter halo. Both models are taken from models in the literature; N35 is the $t=2.875$ Gyr snapshot from the isolated barred $N$-body model [MW] of \cite{stelea24}, and N49 is the $t=8.223$ Gyr snapshot from the satellite merger model M1 of \cite{hunt21,hunt24}. Both models are evolved using the GPU based $N$-body tree code \texttt{Bonsai}\ \rm \citep{bonsai,bonsai-2}, and their full setup is described within their respective papers, but are summarised briefly below.

The initial conditions for N35 were generated using \texttt{Agama}\ \citep{agama} following the MW-like host galaxy with an axisymmetric halo from \cite{vasiliev21}. N35 consists of a $1.2\times10^{10}\ M_\odot$ spherical bulge and a $7.3\times10^{11}\ M_\odot$ dark halo with $8\times10^7$ and $6\times10^8$ particles respectively. The bulge and halo follow \texttt{Agama}'s\rm\ Spheroid potential with $R_\mathrm{s}=0.2,7.0$, $\alpha=1,2$, $\beta=1.8,2.5$ and $\gamma=0,1$ for the bulge and halo respectively. The disc follows an exponential profile and contains $3.2\times10^8$ particles with a total mass of $5\times10^{10}\ M_\odot$, with scale length $R_\mathrm{d}=3$ kpc, scale height $h_\mathrm{d}=0.4$ kpc and $\sigma_{R_0}=90$ \kmsnospace. As described in \cite{stelea24}, this creates a disc which is warmer than the MW. The growth and evolution of the bar in this model are explored in detail in \citet{hunt26}.

The initial conditions for model N49's host galaxy were based on the MWb model from \cite{widrow05} with total mass $\sim6\times10^{11}$ $M_{\odot}$, consisting of a $8.8\times10^8$ particle live NFW halo \citep{navarro97} with $a_{\mathrm{h}}=8.818$ kpc, a $2.2\times10^7$ particle Hernquist bulge \citep{hernquist90} with $a_{\mathrm{b}}=0.884$ kpc and a $2.2\times10^8$ particle exponential disc with $R_\mathrm{d}=2.82$ kpc and $h_{\mathrm{d}}=0.439$ kpc. The simulation follows the merger of the L2 Sagittarius dwarf like satellite taken from \cite{laporte18} into the above host (with a mass ratio of approximately 1:10), and the bar in this model is tidally excited \citep[this galaxy does not form a bar without the interaction; see][]{hunt21}, in contrast with N35.

Table \ref{tabl:simu} provides a summary of the main characteristics of the four MW-like simulations. The 
$N$-body simulations are downsampled prior to constructing the mock catalogues for two reasons: (1) to approximately match the number of stars in the bar region observed in the data (Table~\ref{tabl:data}); and (2) to ensure that all four simulations contain a similar number of particles, making the resulting density maps directly comparable at first glance. We have verified, however, that our results are robust against any reasonable reduction in resolution applied to the MW–like simulations.

\begingroup
\setlength{\tabcolsep}{10pt}
\renewcommand{\arraystretch}{1.5}
\begin{table*}
\caption{Number of sources in the various MW-like simulations before ($N$) and after conversion into \gaia DR3 and \gaianir L10 mock catalogues ($N_\text{mock}$), along with the number of stars located inside the bar region ($N_\text{mock,bar}$). The table compares the true bar pattern speed ($\Omega_\text{p}$), obtained using finite differences, with that derived from the DSS method ($\Omega_\text{p,DSS}$) applied to the mock catalogues. The bar length ($R_{\text{bar}}$) is obtained via the DSS method, namely $R_1$, from the simulations when no observational uncertainties are applied.}
\centering
\begin{tabular}{lccccccccc}
\hline \hline
  &  &  & & \multicolumn{3}{c}{\gaia DR3} & \multicolumn{3}{c}{\gaianir L10} \\
\cmidrule(lr){5-7} \cmidrule(lr){8-10}
 &  $\Omega_\text{p}$ & $R_{\text{bar}}$  & $N$  & $N_\text{mock}$ & $N_\text{mock,bar}$ & $\Omega_\text{p,DSS}$ & $N_\text{mock}$ & $N_\text{mock,bar}$  & $\Omega_\text{p,DSS}$ \\ \hline
TP40  &  40.0  &  3.3 & 24~957~763   &  1~473~523  &  70~457  &  52.4 $\pm$ 0.4   &  5~253~449  &  884~042    &  45.1 $\pm$ 0.2    \\ 
TP42   & 42.0   &  3.3  &  24~957~429   & 1~470~176  &  66~776    &  55.2 $\pm$ 0.5   &  5~212~513  &  854~299    &  47.6 $\pm$ 0.2    \\
N35   &  35.1  &  4.2 & 25~000~000   & 1~531~811  &  142~685    &  51.8 $\pm$ 0.4   &  6~686~327  &  1~823~360    &  39.2 $\pm$ 0.2    \\ 
N49   &  48.6  &  3.3 & 21~974~720   & 1~272~208  &  81~774   &  60.6 $\pm$ 0.4  &  5~154~969  &  1~142~625    &  53.6 $\pm$ 0.1     \\ \hline
\end{tabular}
\tablefoot{The bar pattern speed $\Omega_p$ is provided in \kmskpcnospace, the bar length $R_{\text{bar}}$ in kpc, and the DSS bar region is defined by [$R_0, R_1$] = [$1.0, 3.0$ kpc] for all mocks.}
\label{tabl:simu}
\end{table*}
\endgroup

\section{Methods}
\label{sec:methodology}

In this Section, we describe how we generate \gaia DR3 RGB mock catalogues from the MW-like simulations (Sect.~\ref{subsec:mock}). We also provide a brief overview of how the MW bar pattern speed is estimated in this work, using the DSS method (Sect.~\ref{subsec:dehnen}).

\subsection{\gaia DR3 RGB mock catalogues}
\label{subsec:mock}

As a first step, we oriented all four MW-like simulations such that the Galactic bar is positioned 20$^\circ$ from the Sun–Galactic centre line. Then, to derive \gaia DR3 observable properties for the stellar particles of the MW-like simulations, we used the mock catalogue tool presented in Sect.~5.3 of \citet{scholch25}\footnote{https://github.com/mschoelch24/MockCatalogue}. We implemented the 3D extinction models from \citet{marshall06} and \citet{lallement22}, also used in the \gaia Object Generator (GOG; \citealt{luri14}). To reproduce the photometric properties of the RGB sample, we applied a kernel density estimate (KDE) to the distribution of absolute magnitudes $M_G$ and observed colours $G_\text{BP}-G_\text{RP}$ of the \citetalias{andrae23} sample. From this KDE and the extinction map, we computed an apparent RGB magnitude for each particle in the simulations. These apparent magnitudes then allowed us to assign \gaia DR3 astrometric uncertainties, adopting the prescriptions from the \gaia science performance documentation \citep{pygaia,gaiadr3summary}.

Conversely, the uncertainties in the line-of-sight velocities are neglected in this analysis. This approximation is justified because, for \gaia DR3 (and the expected performance of \gaia DR4, DR5, and \gaianirnospace; see Sect.~\ref{sec:future}), line-of-sight velocity uncertainties are typically small (of order a few km s$^{-1}$) compared to dominant sources of error such as parallax and proper motion uncertainties. Consequently, their contribution to the total uncertainty budget in derived kinematic quantities is subdominant, particularly for analyses primarily sensitive to transverse motions. Neglecting these uncertainties therefore does not affect the conclusions of this work.

After assigning \gaia DR3 astrometric, photometric, and line-of-sight velocity observables to each particle, we apply the \gaia selection function \textit{S(l,b,G)} to predict the number of stars observable in a given pixel at a given magnitude. For simplicity, we simulate the selection function for RC stars, which are considered standard-candle-like. This approximation is justified because the \gaia selection function is primarily driven by apparent magnitude, color, and sky position, and RC and RGB stars occupy largely overlapping regions in these observables. As a result, their completeness is expected to be very similar, with any residual differences due to evolutionary phase being subdominant compared to magnitude- and position-dependent effects. Following \cite{khanna25rc}, we compute the selection function over a fine cylindrical grid ($R,\phi,Z$). For a pixel $i$ at a given line-of-sight ($l,b$) we can compute the maximum observable distance modulus as  
\begin{equation}
\label{eqn:mumax_dust}
    \mu_{\text{max},i} = m_{\lambda,\text{lim}} - M_{\lambda} - A_{\lambda} (l,b,d)_{i},
\end{equation} which depends on the survey magnitude limit $m_{\lambda,\text{lim}}$, the extinction at that pixel $A_{\lambda} (l,b,d)_{i}$, and the absolute magnitude of the RC for which we adopt a \textit{Gaussian} luminosity function $LF_{RC} = \mathcal{N}(\overline{M_{G}}\mathrm{=0.44},\sigma_{\bar{M_{G}}}\mathrm{=0.20})$ following \citet{Hawkins:2017}. The first layer of the selection function, the fraction of observable pixels  $P(\mu_{i} < \mu_{\text{max},i} \mid LF_{RC})$ can then be computed for a range of magnitude limits, in particular for this layer we choose $m_{\lambda,\text{lim}}=G<18$ to mimic \gaia DR3 RC. Next, we consider the probability of a source being actually observed by \gaia and ending up in the subsets of interest (e.g., RVS), for which we use the \texttt{GaiaUnlimited} package \citep{tcg23_sf,acg_23sf}. Specifically we first use the \textit{DR3SelectionFunctionTCG} class to estimate the probability of a source ending up in the entire \gaia DR3 catalogue $S_{\text{top}}(l,b,G)$. Next, using the \textit{SubsampleSelectionFunction} class, we estimate $S_{\text{sub}}(l,b,G)$, the probability of our source ending up in subsets, which is estimated by comparing the number of stars in the catalogue before and after applying selection criteria. Putting it together we have,
\begin{equation}
\label{eqn:rc_sf}
S_{i} = P(\mu_{i} < \mu_{\text{max},i} \mid LF_{RC}) \times S_{\text{top}}(l,b,G) \times S_{\text{sub}}(l,b,G)
\end{equation} where naturally for subsets such as \gaia RC X RVS the magnitude limit would be considerably brighter.

\begin{figure}
    \centering
    \includegraphics[width=\columnwidth]{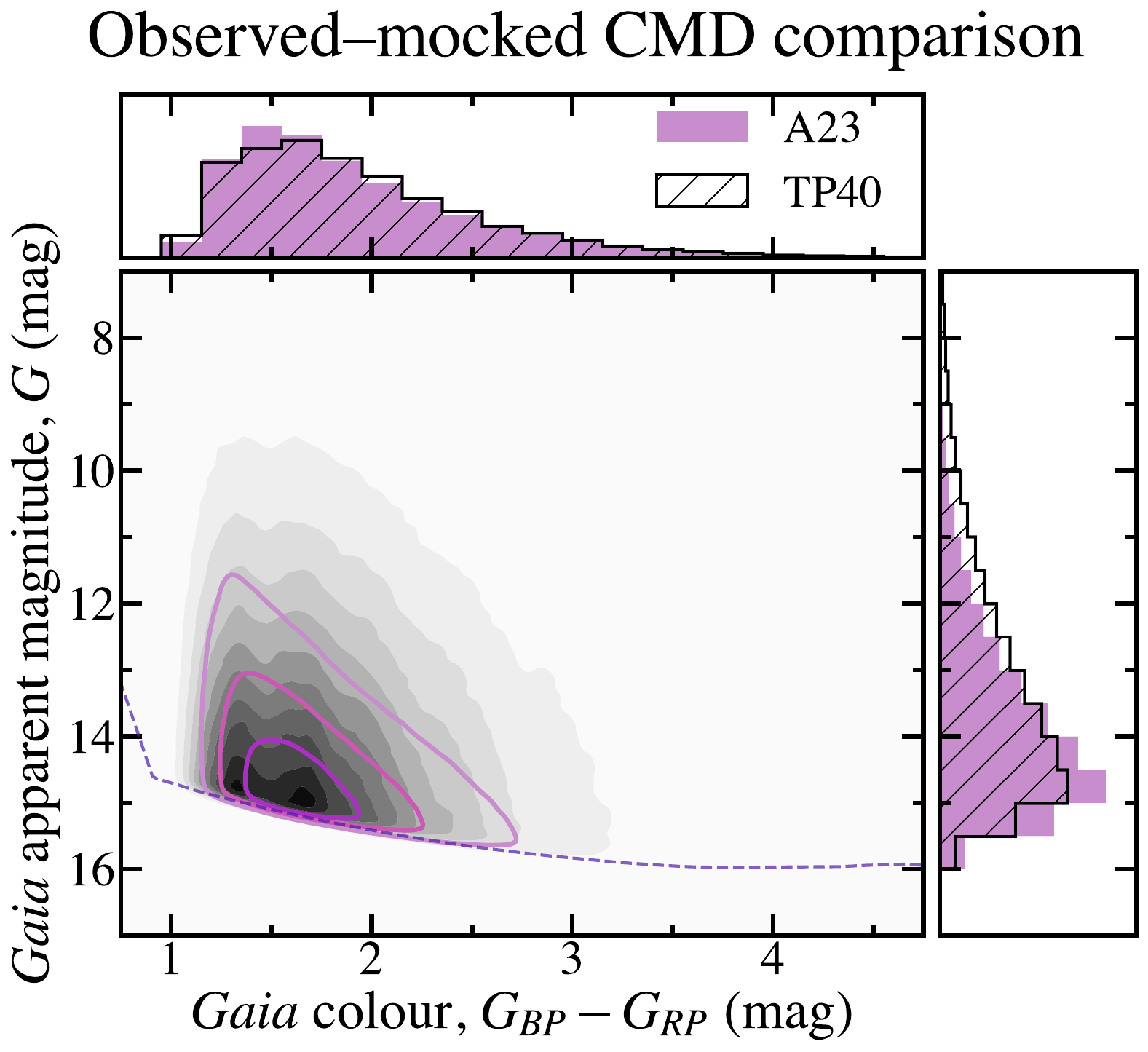}
    \caption{Colour–magnitude diagram (CMD) of the RGB sample from \citetalias{andrae23} (purple contours) compared with the TP40 mock catalogue (black contours). As the mock catalogues from all simulations (TP40, TP42, N35, and N49) differ only slightly, only one representative case is shown. The marginal distributions of the \gaia colour $G_{\text{BP}} - G_{\text{RP}}$ and apparent magnitude $G$ are displayed in the top and right panels, respectively.}
    \label{fig:mock_CMD}
\end{figure}

Finally, we applied the same selection criteria as those used for the RGB \gaia DR3 dataset, prior to deriving the bar pattern speed. In other words, we restricted the sample to stars within $|Z| < 1$~kpc to focus on the Galactic disc, and applied a magnitude cut in $G$ as a function of colour $G_\text{BP}-G_\text{RP}$, which means around $G \sim 15$–$16$~mag (see the purple dashed line in Fig.~\ref{fig:mock_CMD}) produced by the \gaia RVS observational limitations.

\begin{figure*}
    \centering
    \includegraphics[width=0.87\textwidth]{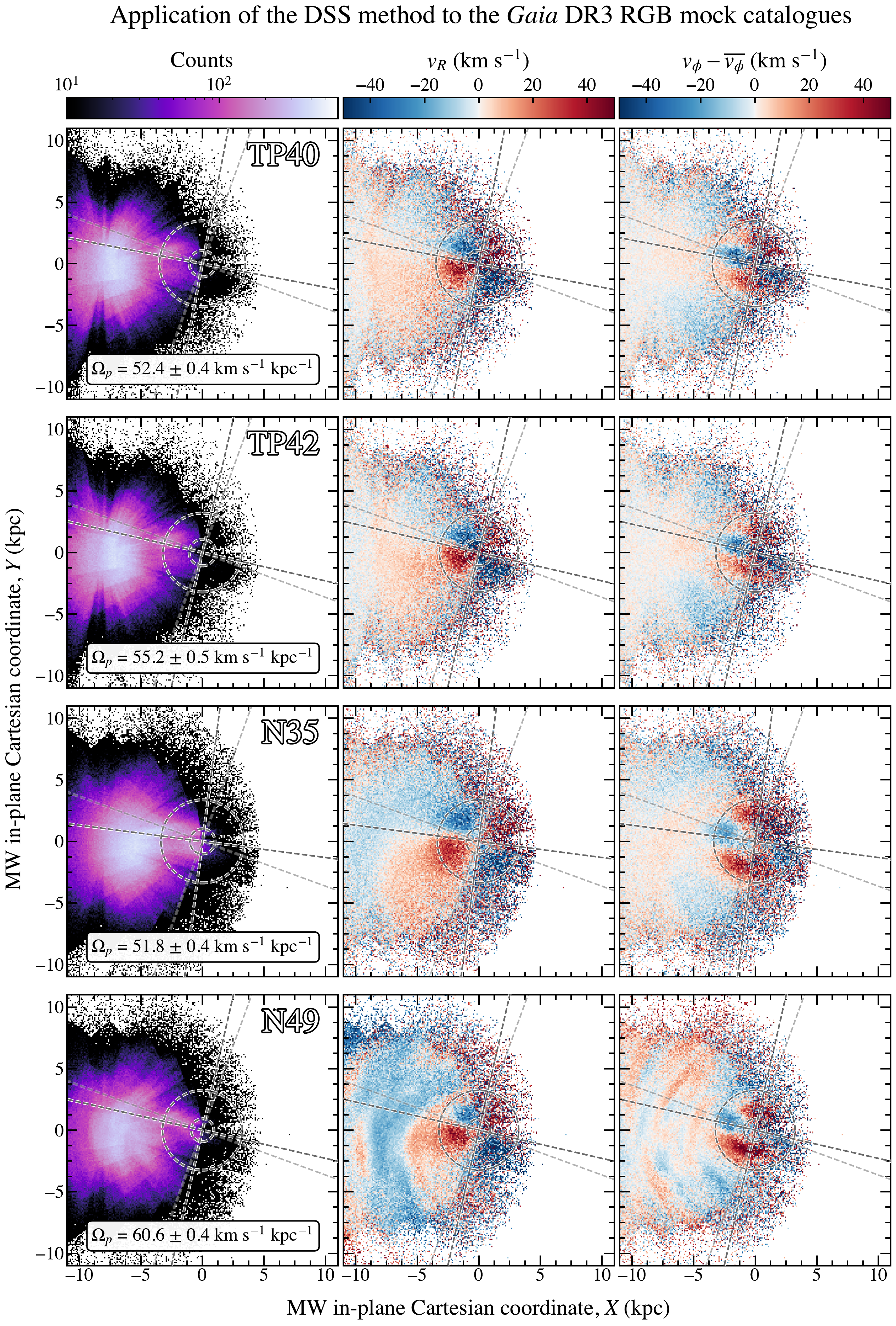}
    \caption{Same as Fig. \ref{fig:data_maps}, but for the MW-like \gaia DR3 RGB mock catalogues. From top to bottom: TP40, TP42, N35, and N49. The transparent light grey dashed lines indicate the input bar’s minor and major axes, oriented at 20$^\circ$ and 110$^\circ$ with respect to the Sun–Galactic center line. In this case, the bar region used in the DSS method corresponds to [$R_0, R_1$] = [$1.0, 3.0$ kpc].}
    \label{fig:mock_maps}
\end{figure*}

Figure~\ref{fig:mock_maps} presents the face-on maps of surface density (left), median radial velocity (centre), and median residual tangential velocity (right) for the MW-like simulation mock catalogues. From top to bottom, the panels correspond to TP40, TP42, N35, and N49. The bar region used in the DSS method is defined as [$R_0, R_1$] = [$1.0, 3.0$]~kpc. As in Fig.~\ref{fig:data_maps}, a decrease in density around $X \sim 0$ is visible, caused by extinction and selection effects in the inner Galaxy, though it appears less pronounced in the mocks. Despite these limitations, the velocity maps reproduce the expected large-scale trends, and the bar’s quadrupole signature remains partially discernible.

\begin{figure}
    \centering
    \includegraphics[width=\columnwidth]{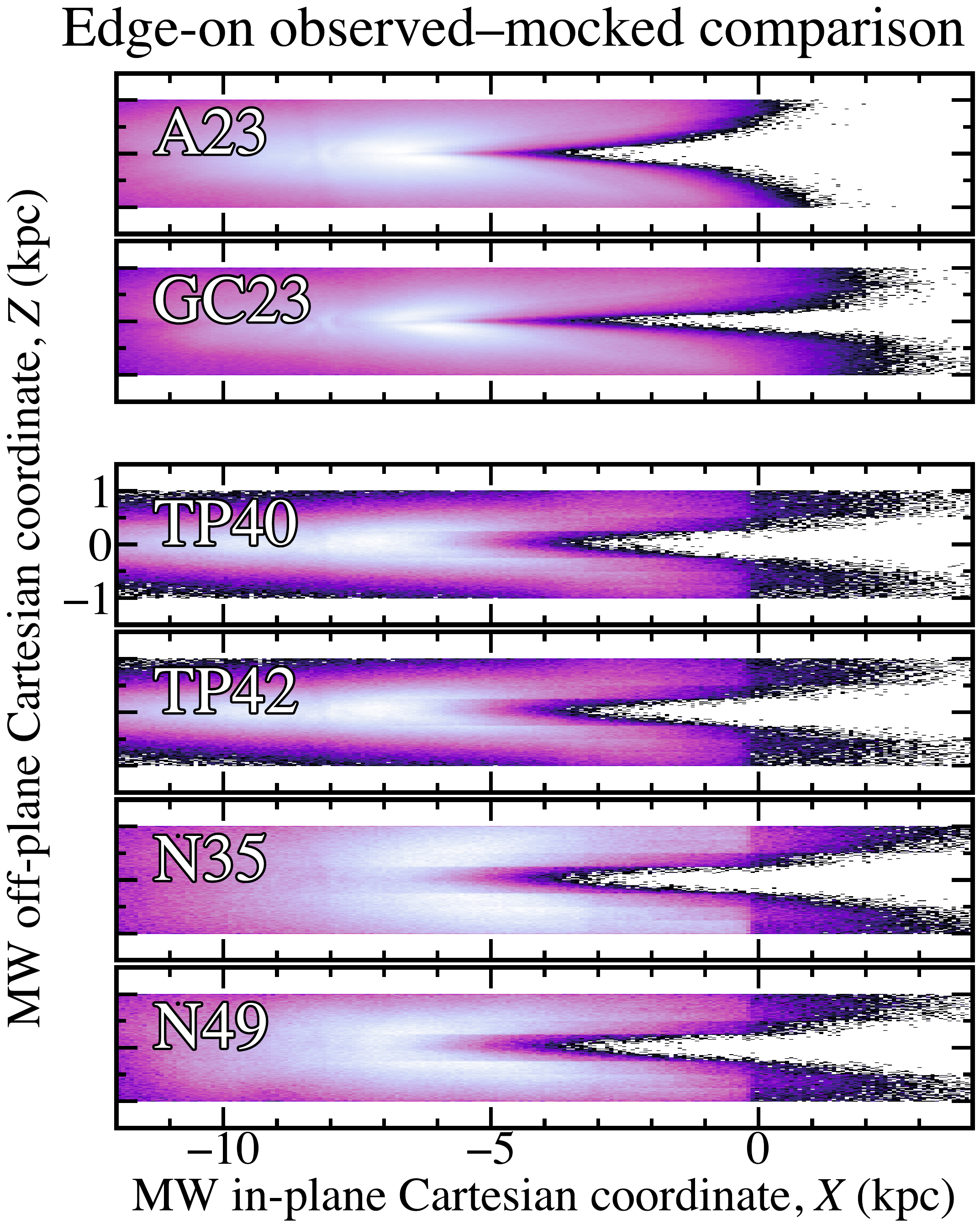}
    \caption{Edge-on view of the \gaia DR3 RGB sample (top two panels) and the mock catalogues from MW-like simulations (bottom four panels). From top to bottom: \citetalias{andrae23}, \citetalias{drimmel22}, TP40, TP42, N35, and N49. In this reference frame, the Sun is located at $(X, Z) \approx (-8.3, 0)$~kpc, while the Galactic center is at $(X, Z) = (0, 0)$~kpc.}
    \label{fig:mock_edgeon}
\end{figure}

Figure~\ref{fig:mock_edgeon} shows the edge-on view of the \gaia DR3 RGB sample (top two panels) and the mock catalogues from the MW-like simulations (bottom four panels). From top to bottom, the panels correspond to \citetalias{andrae23}, \citetalias{drimmel22}, TP40, TP42, N35, and N49. In both the observations and the mock catalogues, the stellar density is nearly zero along the Galactic plane ($Z \sim 0$~kpc) at around $X \sim 5$~kpc, where extinction is too high for \gaia to detect stars in the optical.

Finally, as a last validation of the mock catalogues, Fig.~\ref{fig:mock_CMD} shows the colour-magnitude diagram (CMD) of the RGB sample from \citetalias{andrae23} (purple contours) compared with the TP40 mock catalogue (black contours). The marginal distributions of the \gaia colour $G_{\text{BP}} - G_{\text{RP}}$ and apparent magnitude $G$ are displayed in the top and right panels, respectively. The CMDs of the observed and mock samples show very similar distributions, indicating that the mock catalogues accurately reproduce the \gaia RGB sample. Since the mock catalogues from all simulations (TP40, TP42, N35, and N49) differ only slightly, we show here only the representative case of TP40.

\subsection{Characterisation of the MW bar}
\label{subsec:dehnen}

To estimate the MW bar pattern speed in both observations and mock catalogues, we employ the DSS method. We briefly summarize it here and refer the reader to \citet{dehnen23} for a full description. 

This method, implemented in the publicly available program \texttt{patternSpeed.py}, provides an unbiased, precise, and consistent means to simultaneously measure the bar pattern speed $\Omega_p$ and the bar orientation angle $\phi_b$ from individual snapshots of simulated barred galaxies. It estimates the pattern speed from the time derivative of the $m=2$ Fourier phase, and can be regarded as a refined Fourier-based approach. This method operates under the assumptions that (1) the continuity equation holds, (2) the centre of rotation is known, (3) rotation occurs around the $Z$-axis, and (4) the density is stationary in the rotating frame. The DSS method has been widely applied to both simulations \citep[e.g.][]{bland-hawthorn23,jimenez-arranz24b,semczuk24,jimenez-arranz25b} and 6D observational data of real galaxies \citep{jimenez-arranz24a,zhang24,araya25} to estimate bar pattern speeds.

The DSS method computes the pattern speed within a specified bar region, $[R_0, R_1]$. This region can either be identified automatically — originally by locating the radial interval where the second-order bisymmetric density perturbation reaches a high amplitude with an approximately constant phase angle \citep[see Appendix B of][]{dehnen23} — or supplied directly by the user. In a subsequent refinement, \citet{semczuk24} introduced an updated approach that incorporates higher-order Fourier modes to better constrain the extent of the bar. In this work, we adopt the latter approach because the automatic procedure can be sensitive to noise, incompleteness, and substructure in the data, which may lead to unstable or inconsistent determinations of the bar extent across different samples. For this reason, the DSS method is not ideal for accurately determining the full length of the MW bar. Additionally, the method provides an estimate of the bar angle. For the \gaia DR3 RGB data \citepalias{andrae23,drimmel22}, we define the bar region as $[R_0, R_1] = [1.0, 4.0]\ \mathrm{kpc}$, while for the mock simulations (TP40, TP42, N35, and N49) we adopt $[R_0, R_1] = [1.0, 3.0]\ \mathrm{kpc}$. We have verified that the results presented in this paper are robust against reasonable variations in the adopted bar region.

While large-scale maps of the inner Galaxy, such as those shown in Fig.~\ref{fig:data_maps}, clearly reveal the bar structure, inferring the bar orientation directly from density or kinematic maps is challenging. Distance uncertainties in the inner Galaxy can strongly bias the apparent orientation of both the density enhancement and the associated kinematic signatures \citep[e.g.][]{drimmel22, leung23, hey23, vislosky24}. To obtain a more reliable estimate of the bar angle, and to provide an independent check on the results from the DSS method, we use the bisymmetric velocity model for the tangential velocity $v_\phi$ introduced by \citetalias{drimmel22}. That study showed that the model performs better with $v_\phi$ than with the radial velocity $v_R$, because small heliocentric distance errors have a stronger impact on $v_R$ than on $v_\phi$. Following a similar approach, we measured the phase angle of the second-order perturbation from a Fourier analysis of $v_\phi$ in the four mock catalogues. However, this method cannot be reliably used to determine the bar length and is limited to estimating the bar orientation.

Finally, we also adopt the method of \citet{zhang24} as an independent, kinematic estimate of the MW bar length (see their Sect.~3.4 for details, and our Sect.~\ref{subsec:results_length} for a discussion of the bar length measurement), and additionally investigate its potential for constraining the bar angle. In their study, the authors trace the bar using maps of $\langle |v_R / v_{\rm tot}| \rangle$, since the mean radial velocity fraction provides a robust indicator of the extent of the bar’s $x_1$ orbits \citep{petersen21b}, and therefore of the Galactic bar itself. They show that these maps are less affected by heliocentric distance uncertainties than the raw $v_R$ maps (see the bottom panels of their Fig.~7). While the $v_R$ maps display sign changes along the Sun–Galactic centre line, the $\langle |v_R / v_{\rm tot}| \rangle$ maps better preserve the true orientation of the bar. The authors claim that this robustness arises because distance errors affect both the numerator and denominator of the ratio, leading to an approximate cancellation of their impact. Consequently, the signal in the measured $\langle |v_R / v_{\rm tot}| \rangle$ map can be used to estimate the bar angle. 

In the bottom panels of their Fig.~8, the authors show $\langle |v_R / v_{\rm tot}| \rangle$ as a function of galactocentric radius measured along the bar major axis. Following \citet{petersen21b}, the bar length is inferred from an empirical kinematic criterion based on the relative strength of radial motions, where a characteristic value of $\langle |v_R / v_{\rm tot}| \rangle \simeq 0.3$ was found to correspond to the radius associated with bar-supporting orbital structures and is therefore adopted as a proxy for the bar length.

\section{The Galactic bar in \gaia DR3}
\label{sec:results}

This section presents the MW bar characterisation based on the \gaia DR3 RGB data (Sect.~\ref{sec:data}), with systematic biases quantified using mock catalogues (Sect.~\ref{subsec:mock}). Sections \ref{subsec:results_pattern}, \ref{subsec:results_angle}, and \ref{subsec:results_length} present the inferred bar pattern speed $\Omega_p$, the resulting bar angle $\phi_b$, and the constraints on the bar length $R_b$, respectively.

\subsection{The MW bar pattern speed $\Omega_p$}
\label{subsec:results_pattern}

When applying the DSS method to the \gaia DR3 RGB samples of \citetalias{andrae23} and \citetalias{drimmel22}, we infer a bar pattern speed of $\Omega_p = 43.7 \pm 0.1$ \kmskpc for both datasets (see Fig.~\ref{fig:data_maps}). To quantify the impact of observational effects, such as distance uncertainties, extinction-related systematics, and selection function biases, we apply the same method to the \gaia DR3 mock catalogues with known input bar pattern speeds (see Fig.~\ref{fig:mock_maps}). In all cases, the true pattern speeds are not recovered: the inferred values are systematically higher than the input ones, with offsets ranging from $+12.0$ \kmskpc\ (N49) to $+16.7$ \kmskpc\ (N35). This systematic behaviour is consistently observed across all four simulations, indicating that, when applied to the \gaia DR3 RGB datasets, the DSS method tends to overestimate the bar pattern speed. Thus, $\Omega_p = 43.7 \pm 0.1$ \kmskpc is adopted as a highly conservative upper limit. A comparison between the true values, computed via finite differences, and those obtained from the mock catalogues using the DSS method is presented in Table~\ref{tabl:simu}.

To account for this systematic offset, we can apply a bias correction of $14.4 \pm 2.3$ \kmskpc to the \gaia DR3 RGB results. This correction assumes that the bias is approximately additive and only weakly dependent on the true pattern speed across different mock realizations, suggesting that it is mainly driven by selection effects, extinction, and/or distance uncertainties. Adopting the mean bias and using the scatter among the mocks as a systematic uncertainty, we derive a corrected MW bar pattern speed of $\Omega_p = 29.3 \pm 2.3$ \kmskpcnospace. This value should be interpreted with caution, as the correction is based on a limited set of mock realizations and assumes that their systematics adequately represent those affecting the \gaia DR3 RGB data. Any mismatch between the mocks and the real data — particularly in selection effects, extinction, or distance uncertainties — could introduce additional biases that are not fully captured by this correction.

An important result of this work is that the systematic offset identified in the mock catalogues underscores the limitations of constraining the MW bar pattern speed from current RGB samples using the DSS method. These limitations arise both from the incompleteness of \gaia DR3 and from methodological assumptions that may not fully capture our partial view of the bar (see Sect.~\ref{subsec:disc_bias} for a discussion of possible biases). In Sect.~\ref{sec:future}, we discuss how future \gaia data releases (DR4 and DR5), with improved astrometry and completeness, as well as the proposed \gaianir mission, could help mitigate these systematics and enable more reliable measurements.

\subsection{The angle of the MW bar $\phi_b$}
\label{subsec:results_angle}

To assess the reliability of the bar angle inferred by the DSS method, we apply it to the mock catalogues. The recovered bar angles for TP40, TP42, N35, and N49 are $\phi_b=11^\circ$, $13^\circ$, $8^\circ$, and $13^\circ$ (grey dashed lines in Fig.~\ref{fig:mock_maps}), respectively, thus significantly different from the input value of $\phi_b=20^\circ$ in all the mock simulations (transparent grey dashed lines in Fig.~\ref{fig:mock_maps}).  This demonstrates that, when applied to \gaia DR3 mock RGB samples, the DSS method does not reliably recover the bar angle, systematically underestimating it. For completeness, the method yields bar angles of $\phi_b=2^\circ$ and $7^\circ$ for the \gaia DR3 RGB samples of \citetalias{andrae23} and \citetalias{drimmel22}, respectively.   These values, however, should not be interpreted as meaningful, given the strong biases revealed by the mock catalogues.

\citetalias{drimmel22} proposed that the phase angle of the second order asymmetry of the tangential velocity at low radius is a proxy for the bar angle, and found an angle of $\phi_b \sim 19\degr$ with \gaia DR3 data.
Here, we  perform Fourier transforms of the mock $v_\phi$ and find that the median phase angle of the second order Fourier mode in the bar region for TP40, TP42, N35, and N49 are $29^\circ$, $25^\circ$, $19^\circ$, and $24^\circ$, respectively. The bisymmetric velocity model thus tends to  overestimate the true bar angle (20$^\circ$) but with a less important absolute difference with the ground-truth than for the DSS method. 

\citet{zhang24} analysed the mean fraction of radial to total velocity $\langle |v_R/v_{\rm tot}| \rangle$ and argued it can be a good indicator of the bar length (see below in Sect.~\ref{subsec:results_length}) as well as a proxy for the bar orientation. The Galactic bar angle they visually derived qualitatively from contours of $\langle |v_R/v_{\rm tot}| \rangle$  is $25\degr$. Therefore, following the exercise  performed above with $v_\phi$,  Fourier transforms of $\langle |v_R/v_{\rm tot}|\rangle$ were applied to our mock catalogues. We find absolute differences between the true bar orientation and the second order Fourier asymmetry in our mock $\langle |v_R/v_{\rm tot}|\rangle$ of $1\degr$ to $7^\circ$, though still with a slight trend of over-estimating the true value.  Applied to the \gaia DR3 data, it gives an angle of $24^\circ$  for the \citetalias{andrae23} sample and  $19^\circ$  for the \citetalias{drimmel22} sample (as measured within the same radial range $R=1.5-2.8$ kpc as in  \citetalias{drimmel22}), thus consistent with the value given in \citet{zhang24}. Our analysis thus aligns well with results of \citet{zhang24}, and we think that studying the orientation of the inner bisymmetry in $\langle |v_R/v_{\rm tot}| \rangle$ gives phase angles as reasonable as those obtained with the modelling of the bisymmetry in $v_\phi$ at low radius.

\subsection{The MW bar length $R_b$}
\label{subsec:results_length}
Bar lengths can be obtained through the analysis of the density of stellar tracers. Among the most powerful   methods to infer bar sizes, such as the DSS method, there are also those 
involving the measurement of the radial extent where the phase angle of the inner bisymmetric  perturbation  of density remains constant \citep[see e.g.][ and references therein]{2024ghosh}.
For any other galaxy than the MW, this exercise is   straightforward, particularly when near-infrared imaging is used. However, this is no more the case from a Galactic perspective, as the exact structure of the stellar density at low Galactic latitude, in the bar region, or in the opposite disc half with respect to the Galactic Center is not known. The 3D density that \gaia measures in the bar direction is strongly biased by selection effects, dust extinction and uncertain distances (Figs.~\ref{fig:data_maps} and~\ref{fig:mock_maps}). This makes  the MW bar size estimation from the stellar density highly difficult.

To the contrary, from the view-point of stellar kinematics,  \citet{zhang24} argued that the radius where
 $\langle |v_R / v_{\rm tot}| \rangle = 0.3$ matches the bar length in their mock MW simulations.  We thus assessed this kinematic bar length estimator using our own mock catalogues.
We find that it does not uniquely recover the input bar length when we assume as the true value the bar length obtained from the DSS method applied to the error-free simulations (see Table~\ref{tabl:simu}). The true bar length is underestimated by 1 kpc (i.e. $24\%$) and 0.6 kpc (18\%) for the N35 and N49 mock data, and by 0.3 kpc ($\sim 10\%$) for both test-particles mock data. The correct bar lengths are for $\langle |v_R / v_{\rm tot}| \rangle \sim 0.2$ with the N-body simulated mocks, and $\sim 0.25$ with the test-particles mocks (TP40 and TP42).

These results show that the calibration of this method is somewhat model-dependent.  The difference of $\langle |v_R / v_{\rm tot}| \rangle$ between the test-particle and $N$-body mocks  reflects different gradual transitions in orbital structure near the bar ends, perhaps caused   by the impact of the weak spiral-like features in the $N$-body simulations. 
Given the absence of a robust and uniquely defined threshold, we conclude that adopting $\langle |v_R / v_{\rm tot}| \rangle$ is too uncertain for estimating the bar length, and we therefore choose not to apply this criterion to infer a MW bar length for the \gaia DR3 RGB samples.

\section{Robustness and context of our MW bar characterisation}
\label{sec:discussion}

This section discusses the robustness and puts in context our MW bar characterisation using RGB from the \gaia DR3 dataset. Section \ref{subsec:disc_bias} addresses potential sources of bias affecting our analysis, while Sect.~\ref{subsec:disc_bar_patt} compares our inferred bar properties with previous measurements from the literature.

\subsection{Potential biases on our MW bar characterisation}
\label{subsec:disc_bias}

A potential source of bias in our results is the limited observational coverage on the far side of the bar ($X > 0$), a limitation inherent to any study of the MW bar. To assess whether this affects the determination of the bar properties using the DSS method, we applied the method to simulations without observational errors but with no data for $X > 0$. We also verified that the results remain insensitive to reasonable variations in the input bar length $R_1$ adopted in the DSS method. In all cases, the recovered pattern speed shows no significant variation. Additionally, we experimented with “mirroring” the dataset, filling the unobserved region with a reflection of the data from $X<0$. This procedure did not provide additional information to the DSS method and did not alter the inferred bar pattern speed.

We also investigated whether the input bar angle ($\phi_b=20^\circ$) in our mock affects the inferred bar pattern speed. By testing configurations up to 45$^\circ$, we found that the recovered pattern speed is insensitive to the bar angle. This result reinforces the robustness of the systematic offset and justifies the bias correction of $14.4 \pm 2.3$ \kmskpc applied to the \gaia DR3 RGB results, given that the true MW bar angle may be considered to remain uncertain.

The effects of observational uncertainties are not explicitly assessed in this section. These effects are instead investigated in Sect.~\ref{sec:future}, where we employ mock catalogues constructed to replicate forthcoming \gaia data releases and \gaianir observations.

\subsection{MW bar in context: comparison with previous studies}
\label{subsec:disc_bar_patt}

Recent estimates of the MW bar pattern speed in the \gaia DR3 era \citep[see Fig.~10 of][and references therein]{hunt_vasiliev25} generally place $\Omega_p$ between 30 and 40 \kmskpcnospace. Our analysis, after applying a systematic bias correction, gives $\Omega_p = 29.3 \pm 2.3$ \kmskpcnospace. The offset correction is large compared to the typical literature range: without it, our result ($\Omega_p = 43.7 \pm 0.1$ \kmskpcnospace) would lie above the commonly reported values, and the correction shifts it to the lower end. Our value is numerically close to the recent measurement of $\Omega_p = 24 \pm 3$ \kmskpc reported by \citet[][]{horta25b}; the ranges of the two estimates just touch at $\sim$27 \kmskpcnospace, indicating marginal consistency at the extremes. On the other hand, it is important to keep in mind that \citet{araya25} have shown that different stellar populations can exhibit distinct bar pattern speeds. This highlights the need to consider which tracers are used when comparing results across different studies, as variations in stellar populations can influence the inferred bar dynamics.

We find that  the bisymmetric method applied to $v_\phi$ or the absolute radial-to-total velocity fraction recovers the bar angle in our mock catalogues, with a slight trend of overestimating it. Based on this analysis, the range $\phi_b = 19-24^\circ$ we infer from \gaia\ DR3 data with these methods should encompass the true orientation of the Galactic bar. This agrees well with current estimates of the MW bar angle of $\phi_b \sim 25^\circ \pm 5^\circ$ \citep[see Sect.~4.3.1 of][and references therein]{hunt_vasiliev25}.

Finally, this work did not estimate the bar length from the observations. For reference, recent studies report a range of values: \citet{lucey23} find a dynamical length of 3.5 kpc and a visual extent of ~5 kpc, while \citet{vislosky24} show that the \gaia DR3 radial velocity field is consistent with either a short bar ($R_b \sim 3.6$ kpc) with a moderately strong spiral, or a long bar ($R_b \sim 5.2$ kpc) with weaker spiral structure. These results highlight the current uncertainties in the MW bar length.

\section{Future prospects for the MW bar: from \gaia DR4 to \gaianir}
\label{sec:future}

Motivated by the previous results, which demonstrated that \gaia DR3 RGB mocks cannot recover the bar properties without systematic errors, we repeat the analysis using mock versions of upcoming datasets to assess the future prospects for studying the MW bar. Specifically, we develop mock catalogues for \gaia DR4, DR5, and \gaianirnospace. 

For the forthcoming \gaia data releases (DR4 and DR5), we adopt the parallax uncertainty relations and apply the scaling factors listed in \textsc{PyGaia} \citep{pygaia} to estimate the uncertainties in positions and proper motions. We model the \gaia DR4 and DR5 selection function following the methodology in Sect.~\ref{subsec:mock}, however making a few assumptions in the absence of data to compute the selection function directly. For DR4 we neglect the $S_{\text{top}}(l,b,G)$ for convenience, as this is generally close to unity except in the very faint magnitude bins. We compute the fraction of observable pixels $P(\mu_{i} < \mu_{\text{max},i} \mid LF_{RC})$ choosing $m_{\lambda,\text{lim}}=G<18$ as before. Then we apply a subset selection function to mimic the future \gaia DR4 RVS sample. However, in the absence of real data, and since we cannot use our previous ratio-based method, we instead apply a correction factor. DR4 is expected to provide roughly three times as many sources with reliable line-of-sight velocities compared to DR3. As a simple approximation, we therefore adopt $S_{\text{sub}}(l,b,G)$ from \gaia DR3 RVS and, in each voxel, scale it by a random correction factor drawn uniformly between 1 and 3. For DR5 instead we simplify further by assuming a tentative magnitude limit of $m_{\lambda,\text{lim}}=G<18$ for the RC-like stars with line-of-sight velocities.

For \gaianirnospace, astrometric uncertainties were computed for two proposed mission configurations, a medium (M) and a large (L) design, both assuming both a 5-year (M5 and L5) and 10-year (M10 and L10) mission duration. The uncertainty calculations were performed using the framework of Hobbs et al. (in prep.) which is also used in \citet{scholch25} and Henum et al. (in prep.), adopting a K0III stellar spectrum to model RC stars. As for the \gaia selection function (see Sect.~\ref{subsec:mock}), this approximation for RGB stars is motivated by the fact that RC stars, although a specific subset of the RGB population, have spectral characteristics similar to K-type giants in the \gaia passbands. RC stars are typically late G-type to early K-type giants \citep[approximately G8III–K2III,][]{ruiz-dern2018}, making a K0III giant a suitable proxy for estimating the astrometric uncertainties. To model the selection function for \gaianir RVS we adopt a conservative $m_{\lambda,\text{lim}}=G<18$.

\begin{figure}
    \centering
    \includegraphics[width=\columnwidth]{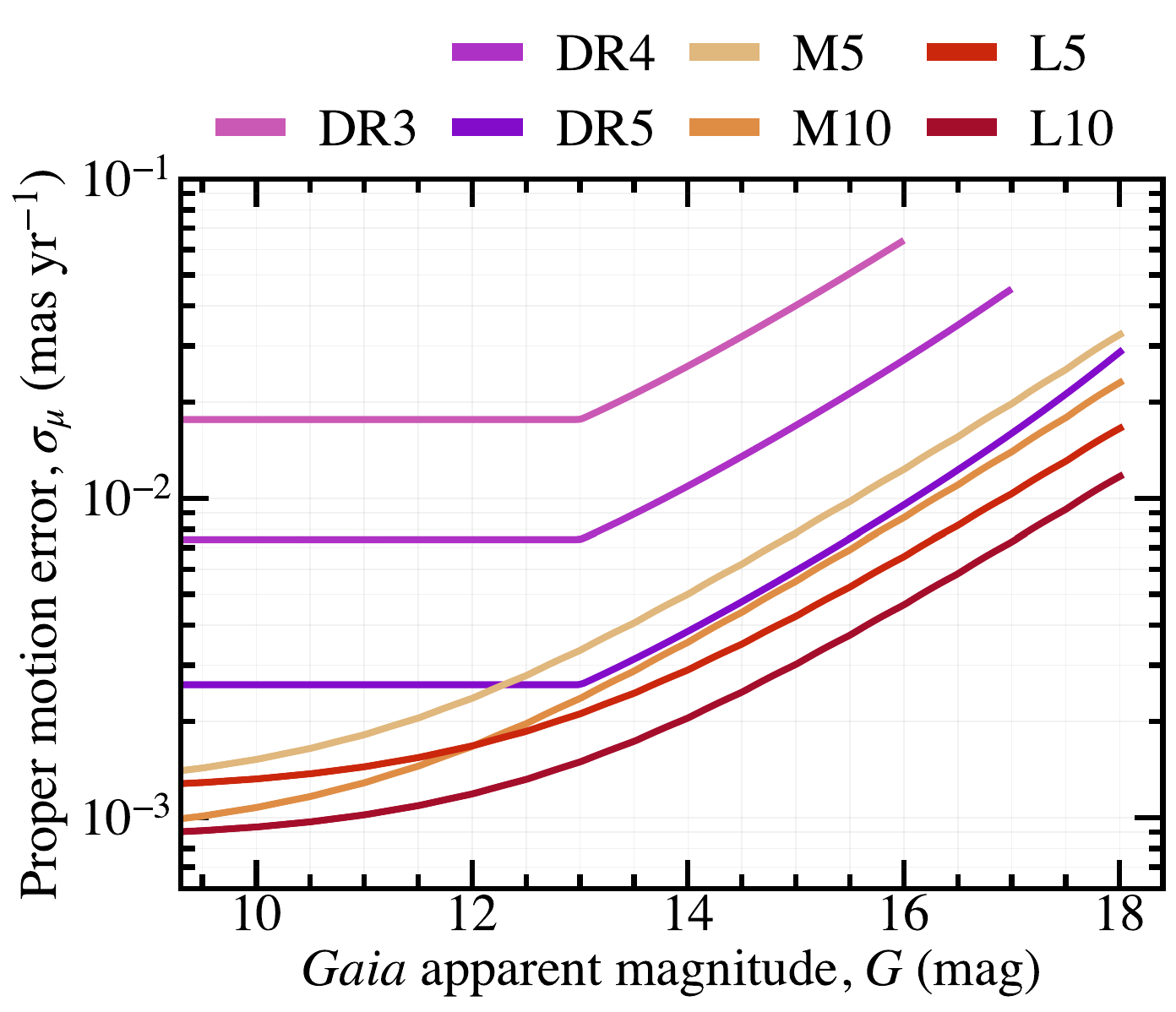}
    \caption{Proper motion uncertainty (solid lines) as a function of the \gaia $G$ magnitude for mock catalogues with RGB error models corresponding to different \gaia releases (DR4 and DR5, shown in purple colours) and \gaianir (M5, M10, L5, and L10, shown in warm colours) with line-of-sight velocity information.}
    \label{fig:rls_errors}
\end{figure}

Figure \ref{fig:rls_errors} shows the uncertainties in the proper motions of the RGB mock catalogues with line-of-sight velocities as a function of the \gaia $G$ magnitude corresponding to different \gaia releases (DR4 and DR5, shown in purple colours) and \gaianir (M5, M10, L5, and L10, shown in warm colours).  The uncertainties decrease significantly from \gaia DR4 and DR5 to the \gaianir missions, with an additional improvement from the medium (M5 and M10, orange colours) to the large (L5 and L10, red colours) \gaianir mission. The uncertainties in right ascension, $\sigma_{\mu_{\alpha*}}$, are systematically slightly larger than those in declination, $\sigma_{\mu_{\delta}}$, but both exhibit the same overall trends. For simplicity, we summarize the results using the combined proper motion uncertainty, $\sigma_\mu = \sqrt{\sigma_{\mu_{\alpha*}}^2 + \sigma_{\mu_{\delta}}^2}$ (solid lines), assuming zero covariance.

The figure illustrates that successive \gaia data releases yield progressively lower proper motion uncertainties across the magnitude range, reflecting the advantages of an increasingly long time baseline. For instance, at $G \sim 16$, the proper motion error decreases from roughly $0.06$ \masyr in DR3 to approximately $0.03$ and $0.01$ \masyr in DR4 and DR5, respectively. Proposed \gaianir missions would extend this trend even further, achieving smaller uncertainties not only for faint stars ($G \sim 17-18$), where errors drop well below $0.01$ \masyrnospace, but also for bright stars ($G \sim 10-12$), where errors fall below $0.001$ \masyrnospace, exceeding the performance of DR5. This highlights that \gaianir would substantially improve proper motion precision across the entire magnitude range.

Figures \ref{fig:mock_maps_dr4} and \ref{fig:mock_maps_dr5} show  face-on maps of surface density (left), median radial velocity (centre), and median residual tangential velocity (right) for the \gaia DR4 and DR5 MW-like simulation mock catalogues, respectively, as similarly presented in Fig.~\ref{fig:mock_maps} for \gaia DR3. Here we can observe how a longer temporal baseline, corresponding to smaller astrometric errors (see Fig.~\ref{fig:rls_errors}) and a less restrictive selection function, enables future \gaia data releases to probe deeper into the Galaxy and sample a larger fraction of the bar. Although the uncertainties in DR4 and DR5 are expected to be significantly reduced compared to \gaia DR3, the true bar pattern speed is still not fully recovered without systematic offsets using the DSS method. Nevertheless, these systematics decrease substantially, from $\sim+14$ \kmskpc in DR3 to $\sim+ 11$ and $\sim+5$ \kmskpc in DR4 and DR5, respectively.

\begin{figure*}
    \centering
    \includegraphics[width=0.87\textwidth]{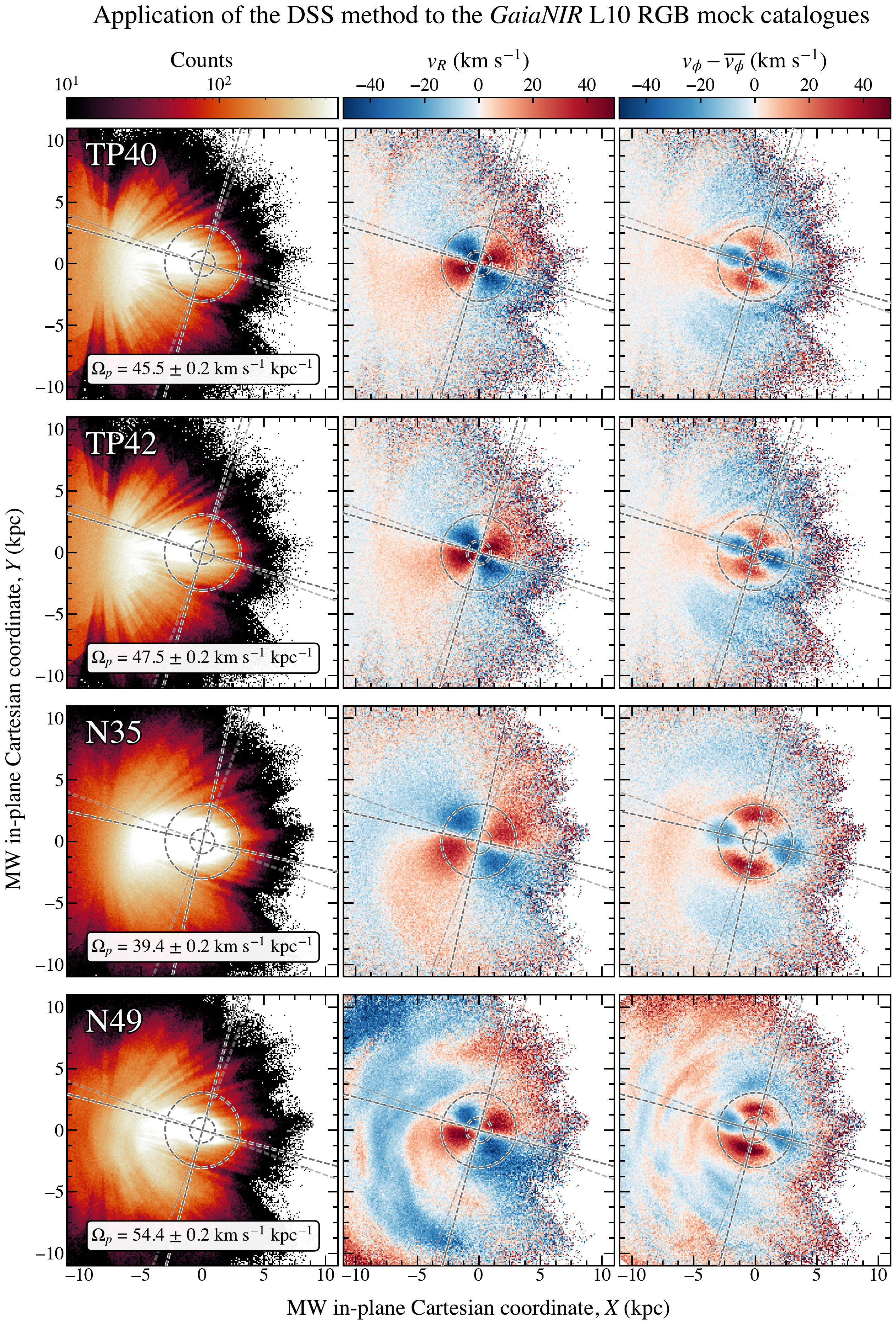}
    \caption{Same as Fig.~\ref{fig:mock_maps} but for \gaianir L10 RGB mock catalogues.}
    \label{fig:mock_maps_gaianir}
\end{figure*}

Figure \ref{fig:mock_maps_gaianir} presents face-on maps of surface density (left), median radial velocity (centre), and median residual tangential velocity (right) for the \gaianir L10 mock catalogues, presented in a manner similar to Fig.~\ref{fig:mock_maps} for \gaia DR3. These maps show a substantial improvement in spatial coverage, reaching further into the inner regions of the MW for a RGB sample with both proper motions and line-of-sight velocities. They also indicate that the bar pattern speed is recovered with a systematic bias of approximately $\sim+5$ \kmskpcnospace, consistent with the \gaia DR5 results. Across all \gaianir mission configurations, the inferred bar pattern speed remains similar. This likely reflects the fact that, at this level, the dominant sources of error are not astrometric precision but factors such as the distance estimation method and/or the selection function. We note, however, that this analysis highlights the significant improvement in astrometric precision at the level of individual stars, as illustrated in Fig.~\ref{fig:rls_errors}. Table~\ref{tabl:simu} summarises the source counts and compares the bar pattern speed inferred using the DSS method for the different \gaianir L10 mock samples. 

Another clear improvement that the new \gaianir mission will enable is on the constraint of the bar angle, as inferred with the three methods used in this present study. With the DSS method, we recover bar angles of $\phi_b=16^\circ$, $16^\circ$, $13^\circ$, and $15^\circ$ for TP40, TP42, N35, and N49, respectively (grey dashed lines in Fig.~\ref{fig:mock_maps_gaianir}). These values are much closer to the input value of $\phi_b=20^\circ$ (transparent grey dashed lines) than those obtained from \gaia DR3 (see Sect.~\ref{subsec:results_angle}). With the method of measuring the phase angle of the   bisymmetry at low radius, we get a bar orientation of $25^\circ$, $25^\circ$, $14^\circ$, and $19^\circ$  ($v_\phi$ data)  and $19^\circ$, $19^\circ$, $14^\circ$, and $17^\circ$ ($\langle |v_R / v_{\rm tot}| \rangle$ data), for TP40, TP42, N35, and N49, respectively, again significantly closer to the real value than for the mock DR3 data.

\section{Conclusions}
\label{sec:conclusions}

The \gaia mission has transformed studies of the MW by delivering precise astrometric measurements for nearly two billion stars, enabling detailed analyses of Galactic structure and dynamics. Nevertheless, a comprehensive dynamical characterisation of the MW bar remains difficult. Fundamental parameters — including the bar’s pattern speed, orientation angle, and length — are still uncertain at the $\sim$20\% level.

In this work, we analyse two large samples of RGB stars with line-of-sight velocity information from \gaia DR3 to explore how incompleteness and distance uncertainties limit current constraints on the MW bar’s properties. Using realistic \gaia DR3 mock catalogues, we assess the sensitivity of astrometry-based inferences to selection effects and systematic distance errors, and provide a critical reassessment of \gaianospace-era measurements of $\Omega_p$. In addition to revisiting previous results, a key goal of this study is to assess how future \gaia data releases (DR4 and DR5) and forthcoming near-infrared astrometry from \gaianir will advance investigations of the Galactic bar. Through mock catalogues, we quantify how the improved completeness, improved astrometry, and more accurate distance estimates from future data releases can mitigate the dominant systematics affecting current measurements. Our main findings and conclusions are as follows:
\begin{itemize}
    \item The DSS method does not reliably recover the bar pattern speed or orientation in \gaia DR3 RGB mock catalogues (Fig.~\ref{fig:mock_maps}), exhibiting a systematic offset of $+14.4 \pm 2.3$ \kmskpcnospace.
    \item When we apply the method to both \gaia DR3 RGB samples (Fig.~\ref{fig:data_maps}), we obtain a bar pattern speed of $\Omega_p = 43.7 \pm 0.1$ \kmskpcnospace, representing a highly conservative upper limit. Incorporating the bias correction derived from the mock catalogues for both samples reduces this estimate to $\Omega_p = 29.3 \pm 2.3$ \kmskpcnospace. This corrected value, however, should be treated with caution due to the limited number of mock realizations and the assumption that their systematics accurately reflect those present in the data.
    \item Bisymmetric perturbations of  $v_\phi$ and  $\langle |v_R / v_{\rm tot}| \rangle$ show phase angles of $\phi_b = 19-24^\circ$ in the bar region.
    \item In our mock catalogues, the method of \citet{zhang24}, based on the radius where $\langle |v_R / v_{\rm tot}| \rangle = 0.3$, fails to reliably recover the bar length.
    \item Future \gaia data releases and \gaianir are expected to improve the determination of the MW bar pattern speed with the DSS method by reducing systematic effects (Figs.~\ref{fig:mock_maps_dr4}, \ref{fig:mock_maps_dr5}, and \ref{fig:rls_errors}) down to $\sim+5$ \kmskpcnospace. 
    \item \gaianir will significantly improve proper motion precision and spatial coverage. At $G \sim 16$, \gaia DR5 achieves $\sim 0.01$ \masyrnospace, while \gaianir will reduce errors to $\sim 0.005$ \masyr for faint stars ($G \sim 17-18$) and below $0.001$ \masyr for bright stars ($G \sim 10-12$), exceeding DR5 performance across the full magnitude range and providing more complete coverage of the Galaxy.
\end{itemize}

Overall, our analysis highlights both the limitations of the current \gaia DR3 and the promise of future \gaia and \gaianir data for refining measurements of the MW bar. Definitive confirmation of these trends, however, will depend on upcoming releases. Alongside enhanced astrometric precision, employing more robust methods will be crucial to minimize systematic biases and accurately determine the bar’s pattern speed, orientation, and structure.

\section*{Data availability}

Both \gaia and \gaianir RGB mock catalogues may be obtained from the corresponding author upon reasonable request.

\begin{acknowledgements}

We are grateful to M. Semczuk for helpful discussions on the behaviour of the DSS method, and for comments that enhanced the clarity of the manuscript.

OJA and DH acknowledges funding from ``Swedish National Space Agency 2023-00154 David Hobbs The GaiaNIR Mission'' and ``Swedish National Space Agency 2023-00137 David Hobbs The Extended Gaia Mission''.

MS and IH acknowledges funding by the European Union under the Horizon Europe Marie Skłodowska-Curie Actions Doctoral Network grant agreement no. 101072454 @HorizonEU research and innovation programme.

SK, RD and EP were supported in part by the Italian Space Agency (ASI) through contract ASI-INAF 2025-10-HH.0 to the National Institute for Astrophysics (INAF).

LC acknowledges  financial support from the French Agence Nationale de la Recherche ANR. 

MRG acknowledges that this work was (partially) supported by the Spanish MICIN/AEI/10.13039/501100011033 and by "ERDF A way of making Europe" by the European Union through grants PID2021-122842OB-C21 and PID2024-157964OB-C21, the Institute of Cosmos Sciences University of Barcelona (ICCUB, Unidad de Excelencia María de Maeztu) through grant CEX2024-001451-M and the project 2021-SGR-00679 GRC de l’Agència de Gestió d’Ajuts Universitaris i de Recerca (Generalitat de Catalunya).

JH acknowledges the support of a UKRI Ernest Rutherford Fellowship ST/Z510245/1.

This work has made use of data from the European Space Agency (ESA) mission {\it Gaia} (\url{https://www.cosmos.esa.int/gaia}), processed by the {\it Gaia} Data Processing and Analysis Consortium (DPAC, \url{https://www.cosmos.esa.int/web/gaia/dpac/consortium}). Funding for the DPAC has been provided by national institutions, in particular the institutions participating in the {\it Gaia} Multilateral Agreement.

\\
\textit{Software:}
\textsc{cmasher} \citep{cmasher},
\textsc{ipython} \citep{ipython}, 
\textsc{jupyter} \citep{jupyter},
\textsc{matplotlib} \citep{matplotlib},
\textsc{mockcatalogue} \citep{scholch25},
\textsc{numpy} \citep{numpy},
\textsc{pandas} \citep{pandas...paper, pandas...software},
\textsc{patternspeed} \citep{dehnen23},
\textsc{pygaia} \citep{pygaia}
\textsc{xgboost} \citep{chen_guestrin16}

\end{acknowledgements}

\bibliographystyle{aa}
\bibliography{mylmcbib}

\begin{appendix}

\section{\gaia DR4 and DR5}
\label{app:mocks_dr4_dr5}

\begin{figure*}
    \centering
    \includegraphics[width=0.87\textwidth]{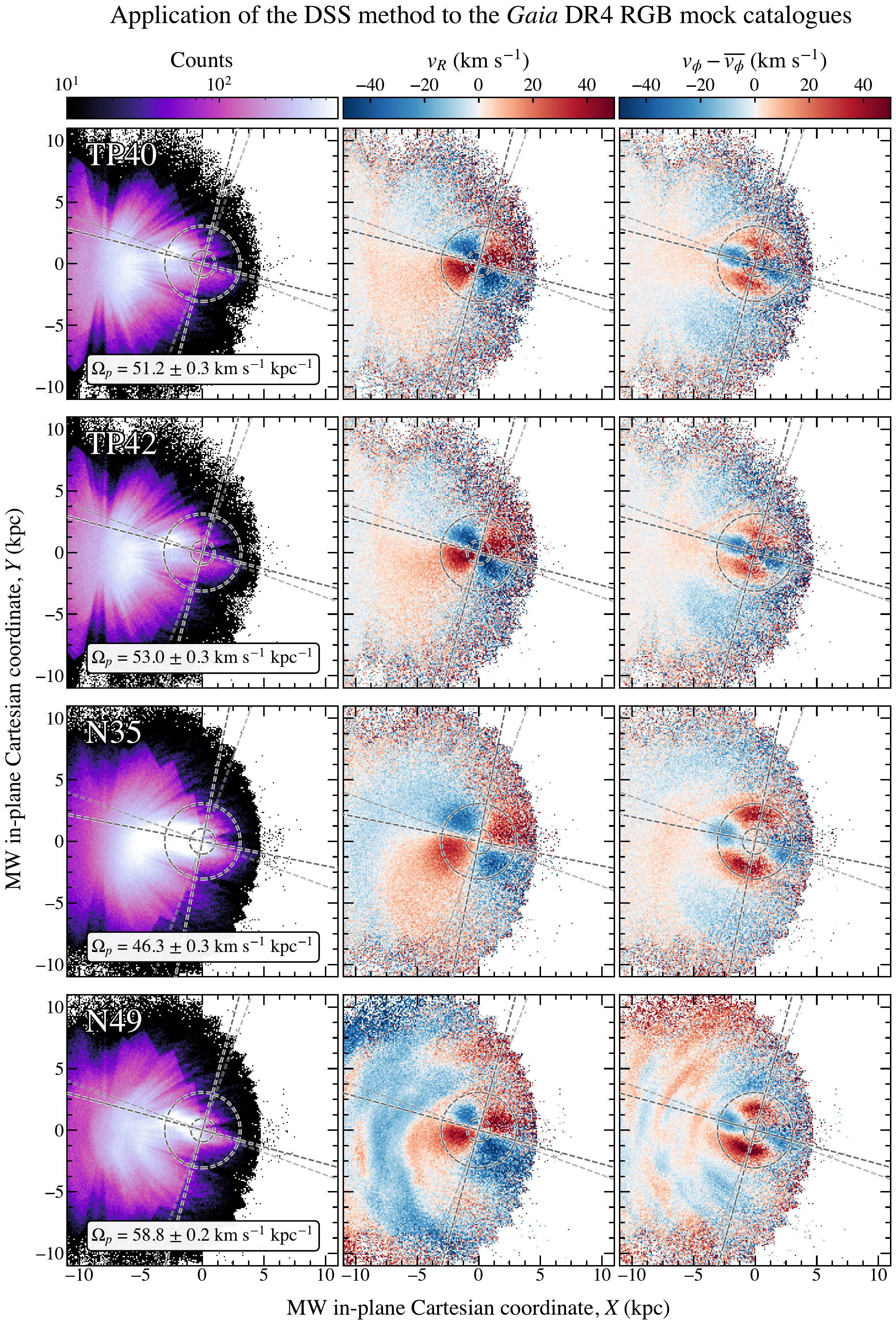}
    \caption{Same as Fig.~\ref{fig:mock_maps} but for \gaia DR4 RGB mock catalogues.}
    \label{fig:mock_maps_dr4}
\end{figure*}

\begin{figure*}
    \centering
    \includegraphics[width=0.87\textwidth]{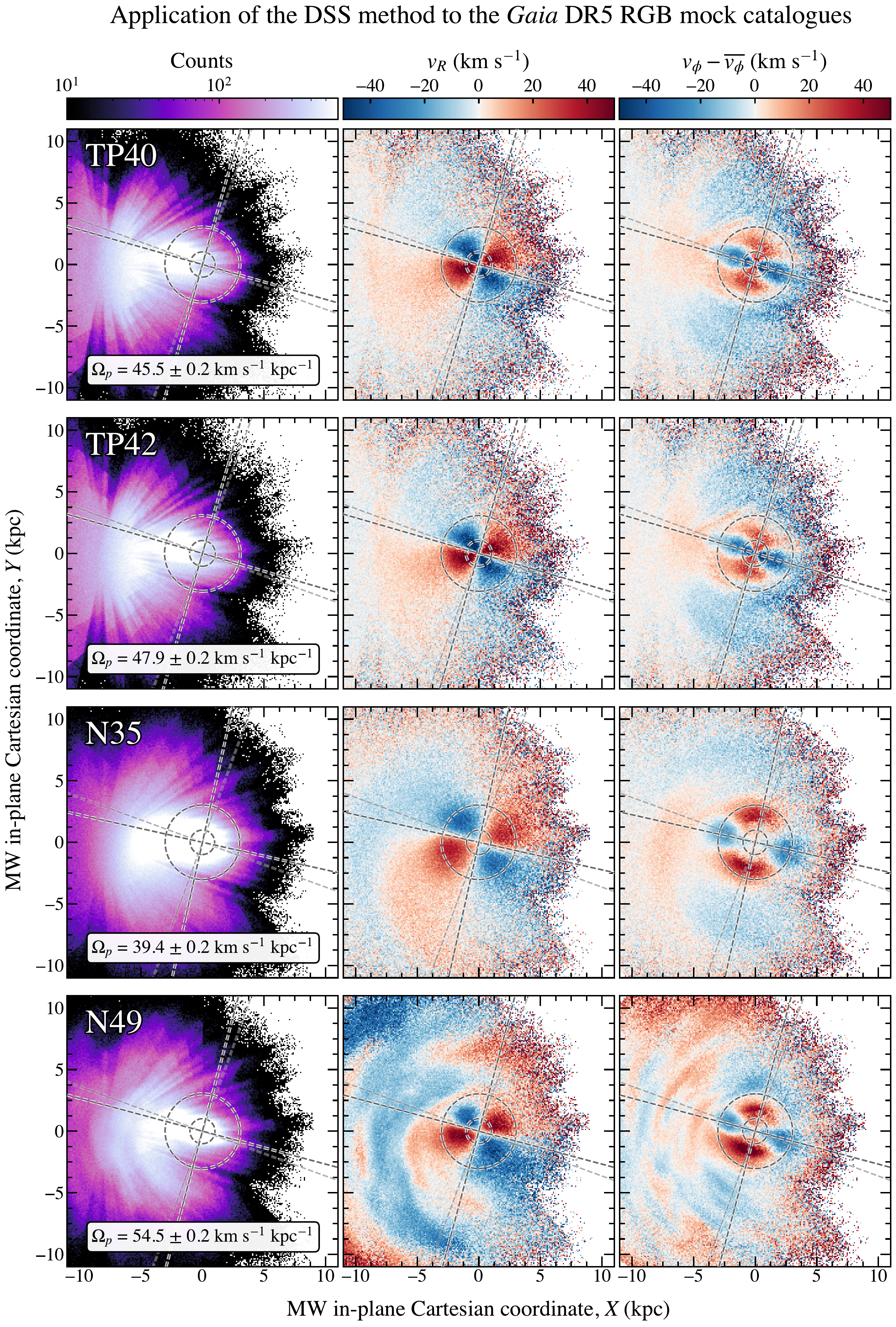}
    \caption{Same as Fig.~\ref{fig:mock_maps} but for \gaia DR5 RGB mock catalogues.}
    \label{fig:mock_maps_dr5}
\end{figure*}

Figures~\ref{fig:mock_maps_dr4} and \ref{fig:mock_maps_dr5} presents the face-on maps of surface density (left), median radial velocity (centre), and median residual tangential velocity (right) for the \gaia DR4 and DR5 MW-like simulation mock catalogues, respectively, as similarly presented in Fig.~\ref{fig:mock_maps} for \gaia DR3. From top to bottom, the panels correspond to TP40, TP42, N35, and N49. The bar region used in the DSS method is defined as [$R_0, R_1$] = [$1.0, 3.0$]~kpc. As in Fig.~\ref{fig:data_maps}, a decrease in density around $X \sim 0$ is visible, caused by extinction and selection effects in the inner Galaxy.

\end{appendix}

\label{lastpage}

\end{document}